# On Fock-Bargmann space, Dirac delta function, Feynman propagator, Angular momentum and SU(3) multiplicity free

**M. Hage-Hassan**
Université Libanaise, Faculté des Sciences Section (1)
Hadath-Beyrouth

**Abstract**

The Dirac delta function and the Feynman propagator of the harmonic oscillator are found by a simple calculation using Fock Bargmann space and the generating function of the harmonic oscillator. With help of the Schwinger generating function of Wigner's D-matrix elements we derive the generating function of spherical harmonics, the quadratic transformations and the generating functions of: the characters of SU (2), Legendre and Gegenbauer polynomials. We also deduce the van der Wearden invariant of 3-j symbols of SU (2). Using the Fock Bargmann space and its complex conjugate we find the integral representations of 3j symbols, function of the series $_2F_1$, and from the properties of $_2F_1$ we deduce a set of generalized hypergeometric functions of SU (2) and from Euler's identity we find Regge symmetry. We find also the integral representation of the 6j symbols. We find the generating function and a new expression of the 3j symbols for SU (3) multiplicity free. Our formula of SU (3) is a product of a constant, 3j symbols of SU (2) by $_3F_2(;1)$. The calculations in this work require only the Gauss integral, well known to undergraduates.

## 1. Introduction

The importance of harmonic oscillator, the group of rotations and SU (3) are well known in quantum physics, nuclear and particles physics. Schwinger [1] in his famous work "on angular momentum "developed the theory in two dimensional harmonic oscillator and use frequently the generating function of this basis. Then, on one side several research groups [2-7] have developed the work of Schwinger to unitary groups and especially SU (3). On the other side the analytic Hilbert space or the Fock Bargmann space, long time known [8], which is derived simply from the generating function of the harmonic oscillator and the spherical harmonics [1,9-14] was studied and applied by Bargmann to the rotation groups [15-16]. Actually the applications of this space are extended to all problems dealing with the harmonic oscillator over to the study of vibrations of the nucleus [17-20].

The calculations are simple using Fock Bargmann space so we propose to extend the applications of this space to the teaching of quantum physics.

Using Dirac transformation, the generating function of the oscillator and the Fock Bargmann space we find the expression of the Dirac delta function and the Feynman propagator. Thus we determine by a simple calculation the Dirac delta function, its properties and the normalization of free particle wave function. Because the techniques of calculation of Feynman kernel function are difficult [21-24], we find by our method the expression of Feynman propagator of the harmonic oscillator by a simple and more coherent method than the well known method using Mehler formula [24-26]. Our method is very useful for the calculation of propagators other than the harmonic oscillator [26].

Using the Schwinger generating function of Wigner's D-matrix [1] we find the generating function of spherical harmonics and also the quadratic transformation. This transformation is very interesting for the determination of the representation (p) of the hydrogen atom [27-28]. We determine also the generating functions: of the characters of SU (2), Legendre polynomials and Gegenbauer polynomials and the spherical basis in terms of creation and destruction operators of harmonic oscillator.

We also observe that the Fock Bargmann space is a subspace of the cylindrical basis of the oscillator and as a result we can use the space of complex conjugate which leads us to the determination of the integral representations (IR), function of Jacobi polynomials, and a class of 3j symbols of SU (n).In the case of SU (2) we express the Jacobi polynomials in term of $_2F_1$ which gives a set of generalized hypergeometric functions $_3F_2$ with argument one and Euler's identity gives Regge symmetry more simply than Vilenkin calculations [29]. We deduce from the (IR) of SU (2) the (IR) of 6j symbols, expression not found in the literature.

Several methods have been proposed to study the unitary groups [3-7] and especially SU (3) but the calculation of 3j symbols is difficult even in the case of multiplicity free [30]. in this work we use the generating functions of the basis of SU (3) [31-32] and the invariant of multiplicity free of this groups then we find the generating function and a new analytical expression of 3j symbols. Our formula of SU (3) is a product of a constant, 3j symbols of SU (2) and $_3F_2$ with argument one.

This paper is organized as follows. Part 2 is a simple revision of the harmonic oscillator and the Fock-Bargmann space. The next section is devoted to the derivation of Dirac delta function. The Feynman kernel of the harmonic oscillator is derived in part 4. In 5 we derive the generating function of spherical harmonics and its applications. In part 6 we derived the van der Wearden invariant of SU (2). The parts seven and eight are devoted to the derivation of integral representations and the symmetry of 3j, 6j symbols. In the last part we find the generating function and the expression of the 3j symbols of SU (3) multiplicity-free.

## 2. Generating function of the oscillator and Fock Bargmann space

In this part we present a brief review of Bargmann Fock space and the generating function of the harmonic oscillator then the oscillator in Dirac notation [9-14].

## 2.1 The generating function of one dimensional harmonic oscillator.

The Schrödinger equation of the harmonic oscillator in one dimension is:

$$H\psi(x) = E\psi(x) \tag{2.1}$$

and
$$H = \frac{1}{2m}(p_x^2 + m^2\omega^2 x^2) \quad [x, p_x] = i\hbar \tag{2.2}$$

Put
$$x = \sqrt{\hbar/(m\omega)}\, q \tag{2.3}$$

we obtain
$$H = \hbar\omega(p^2 + q^2) \tag{2.4}$$

The solution of Schrödinger equation is
$$Hu_n(q) = E_n u_n(q) \text{ and } E_n = \hbar\omega(n + (1/2)). \tag{2.5}$$

With

$$u_n(q) = (\sqrt{\pi} 2^n n!)^{-\frac{1}{2}} e^{-\frac{q^2}{2}} H_n(q) \text{ And } u_n(x) = ((m\omega)/\hbar)^{1/4} u_n(q) \tag{2.6}$$

$H_n(q)$ is the Hermite polynomial with

$$H_n(-q) = (-1)^n H_n(q) \text{ and } u_n(q) = (-1)^n u_n(-q). \tag{2.7}$$

we find then the generating function [12]

$$G(z,q) = \sum_{n=0}^{\infty} \frac{z^n}{\sqrt{n!}} u_n(q) = \pi^{-\frac{1}{4}} \exp(-\frac{q^2}{2} - \frac{z^2}{2} + \sqrt{2}qz) \tag{2.8}$$

In this expression of generating function we note that the functions $\varphi_n(z)$

$$\varphi_n(z) = \frac{z^n}{\sqrt{n!}} \tag{2.9}$$

Form a basis of Analytic Hilbert space that is known by Fock Bargmann space.

And
$$\langle \varphi_n | \varphi_m \rangle = \iint \overline{\varphi_n(z)} \varphi_m(z) d\mu(z) = \delta_{m,n} \tag{2.10}$$

$d\mu(z)$ Is the cylindrical measure

$$d\mu(z) = e^{-(x^2+y^2)} \frac{dxdy}{\pi}, \quad z = x + iy \tag{2.11}$$

we have also
$$f(z) = \int f(z') e^{z\bar{z}'} d\mu(z') \text{ and } e^{\alpha\beta} = \int e^{\alpha\bar{z}} e^{\beta z} d\mu(z) \tag{2.12}$$

## 2.2 Generating function and Dirac notation

We summarize the work of Dirac and we write the relationship between the harmonic oscillator basis and the Fock Bargmann space

### 2.2.1 The basis of harmonic oscillator in Dirac notation

Let
$$a = \frac{1}{\sqrt{2}}(q + \frac{d}{dq}), \quad a^+ = \frac{1}{\sqrt{2}}(q - \frac{d}{dq}) \tag{2.13}$$

We find
$$[a, a^+] = 1, \quad [a, a] = 0, \quad [a^+, a^+] = 0 \tag{2.14}$$

And
$$H = \hbar\omega(N + \frac{1}{2}), \quad N = aa^+ \qquad (2.15)$$

In Dirac notation the basis of the harmonic oscillator becomes

$$|n\rangle = \frac{a^{+n}}{\sqrt{n!}}|0\rangle \qquad (2.16)$$

With $\langle m|n\rangle = \delta_{m,n}$ and $|0\rangle$ is the vacuum state.

**2.2.2 Dirac transformation and generating function**

1- By definition the Dirac transformation is:

$$\langle q|: u_n \to u_n(q) = \langle q|n\rangle \qquad (2.17)$$

$|q\rangle$ is the eigenfunctions of the operator $\hat{q}$ with $\hat{q}|q\rangle = q|q\rangle$.

2- The generating function of the harmonic oscillator is:

$$|z\rangle = e^{za^+}|0\rangle, \quad \langle z'\|z\rangle = e^{\bar{z}'z}$$
And
$$G(z,q) = \langle q|z\rangle \qquad (2.18)$$

The unitary operator is:
$$I = \sum_n |n\rangle\langle n| = \int |z\rangle d\mu(z)\langle z| \qquad (2.19)$$

The generating function $|z\rangle$ is the well known coherent state with:
$$a|z\rangle = z|z\rangle \qquad (2.20)$$

3- The transformation from the representation $\{q\}$ to the representation of the Harmonic oscillator is.

$$\langle q| = \int \langle q\|z\rangle d\mu(z)\langle z| = \int G(z,q)d\mu(z)\langle z| \qquad (2.21)$$

4- The correspondence between the harmonic oscillator and the Bargmann Fock space may be deduced from the relation (2.20):

$$a^+ \to z, \quad a \to \frac{\partial}{\partial z} \qquad (2.22)$$

Because the generators of unitary group may be written in terms of creations and Annihilations operators of the harmonic oscillator [8] then we can use this transformation to express these generators in terms of the variables of Fock Bargmann space.

## 3. The Dirac delta function and the normalization of the free wave

We want to determine the expression $\langle q'|q\rangle$ using the generating function of the harmonic oscillator and the Fock Bargmann space.
We write:

$$\langle q'|q\rangle = \langle q'|I|q\rangle = \sum_{n=0} u_n(q')u_n(q)$$
$$= \int \overline{G(z,q')} G(z,q) d\mu(z) \quad (3.1)$$

.

By replacing $\overline{G(z,q')}$ and $G(z,q)$ by the expression (2.8) we obtain

$$\langle q'|q\rangle = \frac{1}{\sqrt{\pi}} \int \exp[-\frac{q'^2+q^2}{2} - \frac{\bar{z}^2+z^2}{2} + \sqrt{2}(q'\bar{z}+qz)] d\mu(z) \quad (3.2)$$

the arrangement of this expression gives

$$\langle q'|q\rangle = \frac{1}{\pi\sqrt{\pi}} \exp(-\frac{1}{4}(q-q')^2) \int \exp([-2(x-\frac{\sqrt{2}}{4}(q+q'))^2] + \sqrt{2}iy(q-q')) dxdy \quad (3.3)$$

we use the change of variables and after performing the integration we find that

$$\langle q'\|q\rangle = \frac{1}{2\pi} \exp(-\frac{1}{4}(q-q')^2) \int \exp(+ik(q-q')) dk \quad (3.4)$$

Using the Gauss integral we find that

$$\int_{-\infty}^{+\infty} \langle q'\|q\rangle dq' = \frac{1}{2\pi} \int_{-\infty}^{+\infty} \exp[-\frac{1}{4}(q-q')^2 + ik(q-q')] dq' dk = 1 \quad (3.5)$$

But $|q\rangle$ and $|q'\rangle$ are eigenfunctions of the operator $\hat{q}$ then:

$$\langle q'|\hat{q}|q\rangle = q\langle q'\|q\rangle = q'\langle q'\|q\rangle \text{ and } (q-q')\langle q'\|q\rangle = 0 \quad (3.6)$$

it follows that:

$$\exp(-\frac{1}{4}(q-q')^2)\langle q'\|q\rangle = \langle q'\|q\rangle \quad (3.7)$$

Therefore

$$\int_{-\infty}^{+\infty} \langle q'\|q\rangle dq' = \frac{1}{2\pi} \int_{-\infty}^{+\infty} \exp[+ik(q-q')] dq' dk = 1 \quad (3.8)$$

if we use in (3.1) the free wave $Ne^{iqx}$ we find the expression (3.8) if the normalization of the wave function of free particle is: $N = 1/\sqrt{2\pi}$
Finally we write

$$\langle q'\|q\rangle = \delta(q-q') = \frac{1}{2\pi} \int_{-\infty}^{\infty} e^{i(q-q')k} dk \text{ and } \int_{-\infty}^{+\infty} \delta(q-q') dq' = 1 \quad (3.9)$$

We find in a simple and coherent way the Dirac delta function. We deduce also from (3.1) that the delta function is an even function and we obtain the normalization of the wave function of free particle without the help of the distribution theory [12].

## 4. The Feynman propagator of the harmonic oscillator

The Feynman propagator of the oscillator was determined by several methods
The first one is the Feynman path integral, the second is the Schwinger method of Green function, the third is the algebraic method and finally by a direct calculation using the Mehler formula [24-26]. All these methods are difficult for the undergraduates and all the text books gives only the final result. In this section we propose a simple and elementary method for the calculation of this propagator.
The Feynman propagator of the oscillator is:

$$K((x,t),(x',t_0)) = \langle x | e^{-\frac{i}{\hbar}H(t-t_0)} | x' \rangle = \langle x | e^{-\frac{i}{\hbar}H(t-t_0)} I | x' \rangle$$
$$= e^{-i\omega(t-t_0)/2} \sum_n \overline{u_n(x)} e^{-in\omega(t-t_0)} u_n(x') \qquad (4.1)$$

From the orthogonality of the basis $\varphi_n(z)$ and (2. 7-8) we deduce that:

$$K((x,t),(x',t_0)) = \left(\frac{m\omega}{\pi\hbar}\right)^{1/2} e^{-i\omega(t-t_0)/2} \int G(e^{-i\omega(t-t_0)/2}\overline{z},q) G(e^{-i\omega(t-t_0)/2}z,q') d\mu(z) \qquad (4.2)$$

By replacing the expressions under the integral by (2.8) and put $\alpha = \omega(t - t_0)$ we write:

$$\left(\frac{m\omega}{\pi\hbar}\right)^{1/2} e^{-i\alpha/2} \int \exp([-\frac{q^2+q'^2}{2} + e^{-i\frac{\alpha}{2}}(\overline{z}q + zq')\sqrt{2} - \frac{\overline{z}^2+z^2}{2}e^{-i\alpha}] d\mu(z)$$

After arrangement we find that:
$$K((x,t),(x',t_0)) =$$
$$= \frac{1}{\pi}\left(\frac{m\omega}{\pi\hbar}\right)^{1/2} e^{-i\alpha/2} \exp[\frac{i}{2\sin\alpha}[(q^2+q'^2)\cos\alpha - 2xx']] \times E_1 \times E_2 \qquad (4.3)$$

with

$$E_1 = \int \exp[-(2e^{-i\frac{\alpha}{2}}\cos\frac{\alpha}{2})[(x - \frac{\sqrt{2}}{4\cos\frac{\alpha}{2}}(q+q'))^2]] dx \qquad (4.4)$$

and

$$E_2 = \int \exp[-(2ie^{-i\frac{\alpha}{2}}\sin\frac{\alpha}{2})[(y + \frac{\sqrt{2}}{4\sin\frac{\alpha}{2}}(-q+q'))^2]] dy \qquad (4.5)$$

But

$$2e^{-i\frac{\alpha}{2}}\cos\frac{\alpha}{2} = (1+e^{-i\alpha}) \, , \, 2ie^{-i\frac{\alpha}{2}}\sin\frac{\alpha}{2} = (1-e^{-i\alpha}) \text{ and } e^{-i\alpha/2} = \frac{1}{\sqrt{e^{+i\alpha}}},$$

we have also $\qquad q = \sqrt{(m\omega)/\hbar} \, x$ and $\int_{-\infty}^{+\infty} e^{-az^2} dz = \sqrt{\frac{\pi}{a}}$, with Re (a)>0 $\qquad$ (4.6)

Using the above expressions and performing the integration after change of variables we
Find that:

$$E_1 \times E_2 = \pi\sqrt{\frac{1}{2i\sin\alpha}} \quad \text{With} \quad |1\pm\cos\alpha| > 0 \tag{4.7}$$

finally we obtain the expression of Feynman propagator:

$$K((q,t),(q',t_0)) = \sqrt{\frac{m\omega}{2\pi\hbar i\sin\alpha}}\exp[\frac{i}{2\sin\alpha}[(q^2+q'^2)\cos\alpha - 2qq']] \tag{4.8}$$

Consequently we do not encounter the difficulties of the method proposed by Holstein [23] which adopted by all the authors [22-24] and especially the standard books [22]. The same calculation can be used with the generating function [32] to calculate the propagator of the cylindrical basis.

Our method may be applied to the calculation of the partition function [12] and other Feynman propagators [35]. We can also do other calculations with the oscillator representation using the expression (2.21) and Schwinger techniques [1, 24].

## 5. Rotation matrix, generating function of spherical harmonics and other applications

We do a quick review of the work of Schwinger for determining the generating function of the matrix elements of the rotation groups. Then we find the generating function of spherical harmonics, the Kustaanheimo-Steifel transformation-[36-40] and we determine using the Bargmann integral the generating functions: of Gegenbauer, Legendre polynomials and the characters of SU (2). We derive the basis of spherical oscillator in terms of creation and destructions operators using the expansion of the well known free particle wave function.

**5.1 The Fock Bargmann space and the angular momentum**

Schwinger in his work [1] on angular momentum find the generators of SU(2) in terms of creation and destruction operators of two-dimensional harmonic oscillator:

$$\vec{J} = (a_1^+ \quad a_2^+)(\frac{\vec{\sigma}}{2})\begin{pmatrix}a_1\\a_2\end{pmatrix} \tag{5.1}$$

With $\vec{\sigma} = (\sigma_x, \sigma_y, \sigma_z)$ are the Pauli matrices.

We deduce

$$J_+ = J_x + iJ_y = a_1^+ a_2, J_- = J_x - iJ_y = a_2^+ a_1, J_z = (a_1^+ a_1 + a_2^+ a_2)/2 \tag{5.2}$$

Put

$$N = (a_1^+ a_1 + a_2^+ a_2)/2$$

And using the relationships $\vec{J}^2 = J_- J_+ + J_z(J_z+1)$ we find $\vec{J}^2 = N(N+1)$.
The eigenfunctions of $\vec{J}^2$ are the eigenfunctions of N and $J_z$.
We write:

$$|jm\rangle = \frac{a_1^{+\,j+m}a_2^{+\,j-m}}{\sqrt{(j+m)!(j-m)!}}|0,0\rangle \tag{5.3}$$

With
$$j \geq |m|, \quad j = 0, \frac{1}{2}, 1, \frac{3}{2}, 2, \ldots$$

we find the Fock Bargmann basis by applying the transformations (2.22):

$$\varphi_{lm}(u) = \frac{\xi^{l+m}\eta^{l-m}}{\sqrt{(l+m)!(l-m)!}}, u = (\xi, \eta) \tag{5.4}$$

we denote $d\mu(u)$ by the measure of integration

$$d\mu(u) = d\mu(\xi)d\mu(\eta) \tag{5.5}$$

## 5.2 Generating function of the matrix elements of rotation.

The matrix of rotation [42] can be deduced simply from $RR^* = 1$.
Using Euler angles we write:

$$D^j_{(m',m)}(\Omega) = \langle jm'|R|jm\rangle = \langle jm'|e^{-i\psi J_z} e^{-i\theta J_y} e^{-i\varphi J_z}|jm\rangle \tag{5.6}$$

Multiplying by $\varphi_{jm'}(u)\varphi_{jm}(v)r^{2j}$ and after the summation we find the generating function of the matrix elements of rotation

$$\sum_{jmm'} \varphi_{jm'}(u) r^{2j} D^j_{(m',m)}(R) \varphi_{jm}(v) = \exp[{}^t(u)(A)(v)]$$

$$= \exp\left[(\xi \ \eta) \begin{pmatrix} z_1 & -\bar{z}_2 \\ z_2 & \bar{z}_1 \end{pmatrix} \begin{pmatrix} v_1 \\ v_2 \end{pmatrix}\right] \tag{5.7}$$

with

$$z_1 = r\exp(\varphi_1)\cos(\frac{\theta}{2}), \quad z_2 = r\exp(\varphi_2)\sin(\frac{\theta}{2})$$

$$\Omega = (\psi\theta\varphi), \quad \varphi_1 = (\frac{\psi+\varphi}{2}), \quad \varphi_2 = (\frac{\psi-\varphi}{2})$$

and
$$0 \leq r \leq \infty, \ 0 \leq \varphi \leq 2\pi, \ 0 \leq \theta \leq \pi, \ 0 \leq \psi \leq 2\pi$$

It is clear that:

$$\left(\frac{\partial^2}{\partial z_1 \partial \bar{z}_1} + \frac{\partial^2}{\partial z_2 \partial \bar{z}_2}\right) D^j_{(m',m)}(R) = 0 \tag{5.8}$$

Expression may be generalized to SU (n) groups [43-45].

## 5.3 The generating function of spherical harmonics

The relation between the Wigner's D matrix and the spherical harmonics is given by:

$$D^l_{(m,0)}(z_1, \bar{z}_1, z_2, \bar{z}_2) = \sqrt{\frac{4}{(2l+1)}} Y_{lm}(\vec{r}) \tag{5.9}$$

We find from (5.9) that the generating function of spherical harmonics is:

$$\int e^{\lambda(\bar{v}_1\bar{v}_2)} \exp[(\xi \ \ \eta)\begin{pmatrix} z_1 & -\bar{z}_2 \\ z_2 & \bar{z}_1 \end{pmatrix}\begin{pmatrix} v_1 \\ v_2 \end{pmatrix}]d\mu(v_1,v_2) = \exp[\frac{\lambda(\vec{a}\cdot\vec{r})}{2}]$$

$$= \sum_{lm}[\frac{4\pi}{2l+1}]^{\frac{1}{2}}\lambda^l \varphi_{lm}(u)Y_{lm}(\vec{r}) \qquad (5.10)$$

$\vec{a}$ is a vector of length zero, $\vec{a}\cdot\vec{a} = 0$ and has the components

$$a_1 = -\xi^2 + \eta^2, \ a_2 = -i(\xi^2 + \eta^2), \ a_3 = 2\xi\eta \qquad (5.11)$$

we obtain also the quadratic transformation $R^4 \to R^3$:

$$x = z_1\bar{z}_2 + z_2\bar{z}_1, \ y = i(z_1\bar{z}_2 - z_2\bar{z}_1), \ z = z_1\bar{z}_1 - z_2\bar{z}_2 \qquad (5.12)$$

the quadratic transformation $R^4 \to R^3$ [36-40] or Hurwitz transformation [40] has been used first by Kustaanheimo-Steifel in celestial mechanics and was also used by many authors [39] for the connection $R^3$ of hydrogen atom and harmonic oscillator $R^4$. Recently we used it for the derivation of the momentum representation of hydrogen atom [27, 28].

Note: Because there are other quadratic transformations in mathematics and to prevent
 Confusion I proposed Hurwitz transformation [40].

## 5.4 Some applications of the generating function of spherical harmonics

All calculations that we perform in the Fock Bargmann space can be solved with the Gaussian integrals in finite dimensions:

$$(1/\pi^n)\int \prod_{i=1}^n dx_i dy_i \exp(-\bar{z}^t Xz + A^t z + \bar{z}^t \bar{B}) = (\det(X))^{-1}\exp(A^t X^{-1}\bar{B}) \qquad (5.13)$$

With $z = (z_1, z_2, ..., z_n)$

### 5.4.1 Generating function for Legendre polynomials

We put $\xi = \bar{\eta}$ in the formula (5.10) and using (5.13) we find the generating function of Legendre polynomials.

$$\int e^{\vec{r}\cdot\vec{a}/2}d\mu(\eta) = \sum_{l=0}^{\infty}r^l P_l(\cos\theta) = \frac{1}{\sqrt{1-2r\cos\theta+r^2}} \qquad (5.14)$$

### 5.4.2 Generating function of the characters $\chi(R)$ of SU(2).

In the generating function of matrix-D we replace (u) by $(\bar{v})$ we get after integration:

$$\int \exp[^t(\bar{v})(A)(v)]d\mu(v) = \sum_j r^{2j}\sum_m D^j_{(m,m)}(R) = \sum_j r^{2j}\chi(R)$$

$$= 1/(1-2r\cos\frac{\theta}{2}\cos(\varphi+\psi)+r^2) \qquad (5.15)$$

### 5.4.3 The Gaussian integral and the quadratic transformations:

We note that the generalization of the quadratic transformations (5.12) can be written as:

$$z^t A_n z = x_{2n} u + i(\sum_{i=1}^{2n-1} x_i x_i') \tag{5.16}$$

With

$$A_n = \begin{pmatrix} (x_{2n} + ix_{2n-1})I_{2^{n-2}} & A_{n-1} \\ -\bar{A}'_{n-1} & (x_{2n} - ix_{2n-1})I_{2^{n-2}} \end{pmatrix}$$

And $\quad r^2 = \sum_{i=1}^{2n} x_i^2, \ u^2 = \sum_{i=1}^{2n-1} x_i'^2$

$A_n$ Are the Pauli matrices for n = 2 and the Dirac matrices for n = 3.

It is also important to note that we can not deduce these transformations from the Cayley Dixon algebra for n> 3.

We find that the Gauss integral of (5.15) is the generating functions of Gegenbauer polynomials:

$$\int d\mu(z) \exp\{\alpha \bar{z}^t A_n z\} = 1/(1 - 2\alpha x_{2n} - \alpha^2 r^2)^m \tag{5.17}$$

we find also that:

$$A_n = \sum x_i \Gamma_i \text{ and } \Gamma_i^2 = -1, \ \Gamma_i \Gamma_j + \Gamma_j \Gamma_i = \delta_{ij} \tag{5.18}$$

$\Gamma_i$ are elements of Clifford algebra.

It is important to note that the integration with Grassmann variables [33-34] of the formula (5.17) becomes $(1 - 2\alpha x_{2n} - \alpha^2 r^2)^m$. This result can be considered as the extension of the generating function of Gegenbauer polynomials.

### 5.4.4 The spherical basis of the harmonic oscillator

We can deduce from the formula (2.18) and the development of the free wave the basis of three dimensions harmonic oscillator $|nlm\rangle$ in terms of creations and destructions operators. $|nlm\rangle$ Is the eigenfunctions of the operators $N$, $\vec{L}^2$ and $L_z$.
$\vec{L}$ And $L_z$ are the angular momentum and its projection on the z-axis.
We have

$$\exp[z_1 a^+ + z_2 a^+ + z_3 a^+]|0,0,0\rangle = 4\pi \sum_{nlm} j_l(z\rho) Y_{lm}^*(\vec{z}) Y_{lm}(\vec{a}^+)|0,0,0\rangle$$
$$= 2\pi^2 \sum_{nlm} N_{nl}^2 (z\rho)^{2n+l} Y_{lm}^*(\vec{z}) Y_{lm}(\vec{a}^+)|0,0,0\rangle \tag{5.19}$$

and

$$|nlm\rangle = N_{nl} \rho^{+2n} Y_{lm}(\vec{a})|0,0,0\rangle, \rho^2 = a_x^{+2} + a_y^{+2} + a_z^{+2} \tag{5.20}$$

the image of these functions in Fock Bargmann space is:

$$\varphi_{nlm}(z) = N_{nl} z^{2n} Y_{lm}(\vec{z}), z^2 = z_x^2 + z_y^2 + z_z^2 \tag{5.21}$$

The normalization $N_{nl}$ is given by [2]:

$$N_{nl} = (-1)^n \sqrt{4\pi / [(2n + 2l + 1)!!(2n)!!]} \tag{5.22}$$

# 6. Generating function of spherical harmonics and Van der Wearden invariant of 3-j symbols.

Van der Wearden [46] determined the invariant of SU (2), method known to Weyl [48, 49], and deduce the Wigner 3j symbols of this group. We will determine first the Van der Wearden invariant using the generating function of spherical harmonics. Second we derive the generating function of the 3j symbols and finally we give the expression of these symbols or Van der Wearden formula.

We consider the integral

$$I = \int \exp[-\vec{r}^2 + 2(\alpha_1 \vec{a}_1 \cdot \vec{r} + \alpha_2 \vec{a}_2 \cdot \vec{r} + \alpha_3 \vec{a}_3 \cdot \vec{r})]dxdydz \qquad (6.1)$$

With

$$u^1 = (\xi_1, \eta_1),\ u^2 = (\xi_2, \eta_2),\ u^3 = (\xi_3, \eta_3),$$

And

$$\vec{a}_i \cdot \vec{a}_j = -2[u^i u^j]^2 \ avec\ [u^i u^j] = [\xi_i \eta_j - \eta_i \xi_j]. \qquad (6.2)$$

$[u^2 u^3]$, $[u^3 u^1]$ and $[u^1 u^2]$ are the elementary invariants of SU (2).
After integration we obtain:

$$I = \exp[-2\alpha_1\alpha_2 \vec{a}_1.\vec{a}_2 - 2\alpha_1\alpha_3 \vec{a}_1.\vec{a}_3 - 2\alpha_2\alpha_3 \vec{a}_2.\vec{a}_3] \qquad (6.3)$$

the development of the integral (6.1) gives

$$I = \sum_{l_i} 2^{4L} [\prod_{i=1}^{3} (\alpha_i^{l_i} \sqrt{\frac{4\pi}{2l_i+1}})] \times [\int_0^\infty \exp\{-r^2\} r^{L+2} dr]$$
$$\times \left[ \sum_{m_i} \prod_{i=1}^{3} \varphi_{l_i m_i}(u^i) \int \prod_{i=1}^{3} Y_{l_i m_i}(\theta\varphi) \right] \sin\theta d\theta d\varphi \qquad (6.4)$$

With $L = l_1 + l_2 + l_3$

We use the well-known result of the theory of angular momentum

$$\int \prod_{i=1}^{3} Y_{l_i m_i}(\theta\varphi) \sin\theta d\theta d\varphi = \prod_{i=1}^{3} (\frac{(2l_i+1)}{4\pi})]^{\frac{1}{2}} \begin{pmatrix} l_1 & l_2 & l_3 \\ 0 & 0 & 0 \end{pmatrix} \begin{pmatrix} l_1 & l_2 & l_3 \\ m_1 & m_2 & m_3 \end{pmatrix}. \qquad (6.5)$$

1- After the integration of (6.4) and the identification with the second member of the integral we obtain the Van der Wearden invariant of 3-j symbols:

$$\sum_{m_i} \left[ \prod_{i=1}^{3} \varphi_{l_i, m_i}(u^i) \right] \begin{pmatrix} l_1 & l_2 & l_3 \\ m_1 & m_2 & m_3 \end{pmatrix} = \frac{[u^2 u^3]^{(L-2l_1)} [u^3 u^1]^{(L-2l_2)} [u^1 u^2]^{(L-2l_3)}}{\sqrt{(L+1)!(L-2l_1)!(L-2l_2)!(L-2l_3)!}} \qquad (6.6)$$

this expression is valid for integers and half-integers of $l_1, l_2$ and $l_3$ [43]. More this formula is very interesting because we can deduce from it by a simple calculus all the particulars cases of the 3j symbols.

2- We deduce the generating function of 3j symbols of SU (2) from (6.6) by multiplying it by:

$$\varphi_{l_3(l_1l_2)}(\tau) = [(L+1)]^{\frac{1}{2}} \frac{[\tau_3]^{(L-2l_3)}[\tau_2]^{(L-2l_2)}[\tau_1]^{(L-2l_1)}}{\sqrt{(L-2l_3)(L-2l_2)(L-2l_1)}}, \qquad (6.7)$$

After summing with respect to $j_i = l_i = 0, 1/2, 1, ..$ we obtain the Schwinger formula:

$$\sum_{j_i,m_i} \varphi_{j_3(j_1j_2)}(\tau) \left[ \prod_{i=1}^{3} \varphi_{j_i,m_i}(u^i) \right] \begin{pmatrix} j_1 & j_2 & j_3 \\ m_1 & m_2 & m_3 \end{pmatrix} = \exp\left( \begin{vmatrix} \tau_1 & \tau_2 & \tau_3 \\ \xi_1 & \xi_2 & \xi_3 \\ \eta_1 & \eta_2 & \eta_3 \end{vmatrix} \right) \qquad (6.8)$$

the symmetries of the 3j symbols can be deduced from the invariance of the determinant: permutation of columns, permutation of rows and transposition [16].

3- The Van der Wearden formula for 3j symbols can be derived simply form (6.6):

$$\begin{pmatrix} l_1 & l_2 & l \\ m_1 & m_2 & m_3 \end{pmatrix} = (-1)^{2(l_2-l_1)} \Delta(l,m) \frac{(l-l_1+l_2)!(l-m_3)!(l_2-m_2)!}{(-l+l_1+l_2)!(l-l_2+m_1)!(l-l_1-m_2)!} \times$$

$$_3F_2(-l_2-m_2, -l_1+m_1, l-l_1-m_2; l-l_1-m_2+1, l-l_2+m_1+1; 1) \qquad (6.9)$$

With

$$\Delta(l,m) = (-1)^{l_2+m_2} \sqrt{\frac{(l+l_1-l_2)!(-l+l_1+l_2)!(l+m_3)!(l_1-m_1)!}{(l-l_1+l_2)!(l+l_1+l_2+1)!(l-m_3)!(l_1+m_1)!(l_2+m_2)!(l_2-m_2)!}}$$

The method of invariants has been the subject of many studies [50-51] but Resnikoff [51] extended the work of Bargmann to SU (3) groups after the interesting work of Moshinsky using the harmonic oscillator basis [3]. We also note that the generalization of this method for SU (n) for n> 3 is not so easy.

## 7. Integral representations, expressions and symmetries of 3j symbols of SU (2)

We treat first the realization in the dual space of Fock-Bargmann and we deduce the integral representation of SU (2) and the sets of generalized hypergeometric functions for 3j symbols. Then we deduce from Euler identity the Regge symmetry of 3j symbols [52].

### 7.1 The realization in the dual space of Fock-Bargmann

The solution of Schrödinger equation of the cylindrical basis of the oscillator is:

$$\Phi_{jm}(\rho,\alpha) = \frac{1}{\sqrt{\pi}} \sqrt{\frac{(j-|m|)!}{(j+|m|)!}} \exp(-\frac{\rho^2}{2}) L_{j-|m|}^{2|m|}(\rho^2) \rho^{2|m|} e^{-2im\alpha} \qquad (7.1)$$

we find the Fock Bargmann space for $m \geq 0$ and its complex conjugate for $m \leq 0$.

If we consider the Fock Bargmann space as a subspace of the cylindrical basis of the oscillator then the correspondence principle implies that the adjoint operators are:

$$z = x + iy \Rightarrow z^* = x - iy \quad \text{And} \quad \left(\frac{\partial}{\partial z}\right)^* = -\left(\frac{\partial}{\partial x}\right) + i\left(\frac{\partial}{\partial y}\right) = -\left(\frac{\partial}{\partial \bar{z}}\right) \quad (7.2)$$

The generators of unitary groups in the Fock Bargmann space [8] are:

$$E_{ij} = \sum_{ij} z_i (\partial / \partial z_j)$$

Then the adjoint operators are:

$$E_{ij}^* = -\sum_{ij} \bar{z}_i (\partial / \partial \bar{z}_j) \quad \text{And} [E_{ij}, E_{ij}^*] = 0.$$

The generators of the coupling are:

$$\Lambda_{ij} = E_{ij} + E_{ij}^*$$

We will treat the case of SU (2) which is valid for SU (3) [44] and SU (N).
In the case of SU (2) we have the correspondence: $\vec{J} \leftrightarrow \vec{J}^*$ and $[\vec{J}, \vec{J}^*] = 0$.
The eigenfunctions of $\vec{J}^{*2}$ and $J_z^*$ are

$$\varphi_{jm}(-\bar{z}_2, \bar{z}_1) = \frac{\rho^{2j}}{\sqrt{(2j)!}} D^j_{((m,j))}(\Omega) \quad (7.3)$$

the eigenfunctions of $\vec{J}^2$ and $J_z$ are:

$$\varphi_{jm}(z_1, z_2) = \frac{\rho^{2j}}{\sqrt{(2j)!}} D^j_{(m,j)}(\Omega) \quad (7.4)$$

We note that the coupling of two angular momentum result from the coupling of the elements of rotation matrix so we can use the Gaunt integral for the determination of 3j symbols [8].

**7.2 The integral representation of SU (2)**
the Gaunt formula is [42, 43]:

$$\int d\Omega D^{j_1}_{(m_1,m'_1)}(\Omega) D^{j_2}_{(m_2,m'_2)}(\Omega) D^{j_3}_{(m_3,m'_3)}(\Omega) = \begin{pmatrix} j_1 & j_2 & j_3 \\ m'_1 & m'_2 & m'_3 \end{pmatrix} \begin{pmatrix} j_1 & j_2 & j_3 \\ m_1 & m_2 & m_3 \end{pmatrix} \quad (7.5)$$

Put $m'_1 = j_1$, $m'_2 = -j_2$ and $m'_3 = j_2 - j_1$.
We can calculate the particular case using (7.5) or much simpler from (6.6):

$$\begin{pmatrix} j_1 & j_2 & j_3 \\ j_1 & -j_2 & j_2 - j_1 \end{pmatrix} = (-1)^{-2j_1 + 2j_2} \sqrt{\frac{(2j_1)!(2j_2)!}{(j_1 + j_2 + j_3 + 1)!(j_1 + j_2 + j_3)!}} \quad (7.6)$$

we therefore deduce the integral representation of 3j symbols:

$$\begin{pmatrix} j_1 & j_2 & j_3 \\ m_1 & m_2 & m_3 \end{pmatrix} = \frac{(-1)^{2(j_2 - j_1) + j_2 + m_2} \Gamma_3}{\Gamma_2 \Gamma_1 \sqrt{(J - 2j_2)!(J - 2j_1)!}} \int_0^\pi [(\cos\frac{\theta}{2})^{2(j_2 - m_2)+1} (\sin\frac{\theta}{2})^{2(j_1 - m_1)+1}$$
$$\times P^{((m_3 + j_1 - j_2, m_3 - j_1 + j_2))}_{j_3 - m_3}(\cos\theta)] d\theta \quad (7.7)$$

with $\Gamma_i = \sqrt{(j_i - m_i)!(j_i + m_i)!}$

### 7.3 the set of generalized hypergeometric functions for 3j symbols

we will write the familiar Jacobi polynomials [25] in terms of the hypergeometric functions:

$$P_n^{(\alpha,\beta)}(\cos\theta) = \frac{(-1)^\beta \Gamma(n+\alpha+1)}{n!\Gamma(1+\alpha)} {}_2F_1(n+\alpha+\beta+1,-n;1+\beta;\cos^2\frac{\theta}{2}) \qquad (7.8)$$

$$P_n^{(\alpha,\beta)}(\cos\theta) = \binom{n+\alpha}{n} {}_2F_1(n+\alpha+\beta+1,-n;1+\alpha;\sin^2\frac{\theta}{2}) \qquad (7.9)$$

$$P_n^{(\alpha,\beta)}(\cos\theta) = \frac{\Gamma(n+\alpha+1)}{n!\Gamma(1+\alpha)}\left(\cos^2\frac{\theta}{2}\right)^n {}_2F_1(m'-j,-j-m;1+m'-m;-tg^2\frac{\theta}{2}) \qquad (7.10)$$

it is possible for any of finites series to reverse the order of terms, from the end to the beginning. In general, we can write

$$_{A+1}F_B[-a,(b);(c);z] = \frac{((b))_a(-z)^a}{((c))_a} \times$$

$$_{B+1}F_A[-a,1-a-(c);1-a-(b);(-1)^{A+B}/z] \qquad (7.11)$$

With $z = \cos^2(\theta/2), \sin^2(\theta/2)$ and $\cot^2(\theta/2)$.

Using the expressions (7.8)-(7-11)) in (7.7) and perform the integration we get the set of generalized hypergeometric functions for 3j symbols.

### 7.4 Wigner's expression for 3j symbols

We write the Jacobi polynomials in terms of $\cot^2(\theta/2)$ in the expression (7.7) and after integration we get the Wigner's expression for 3j symbols:

$$\begin{pmatrix} j_1 & j_2 & j_3 \\ m_1 & m_2 & m_3 \end{pmatrix} = (-1)^{j_2-j_1+m_3} \Delta(j,m) \frac{(j_1+j_2+m_1)!}{(j_1-j_2+m_3)!} \times$$

$$_3F_2(-j_3+m_3,-j_3+j_1-j_2,j_1-m_1-1;j_1-j_2+m_3+1,-j_3-j_2-m_1;1) \qquad (7.12)$$

### 7.5 Euler's identity and Regge symmetry of SU(2)

To determine the symmetries of the 3j symbols we write in the expression (7.7) the Jacobi polynomials in terms of hypergeometric functions and then we use the Euler identity

$$F(\alpha,\beta;\gamma;z) = (1-z)^{\gamma-\alpha-\beta} F(\gamma-\alpha,\gamma-\beta;\gamma;z) \qquad (7.13)$$

we find

$$\int_0^\pi d\theta(\sin^2(\theta/2))^\rho (\cos^2(\theta/2))^\sigma {}_2F_1(n+\alpha+\beta+1,-n;1+\beta;\cos^2(\theta/2)) =$$

$$\int_0^\pi d\theta(\sin^2(\theta/2))^{\rho-\alpha}(\cos^2(\theta/2))^\sigma {}_2F_1(-n-\alpha,1+\beta+n;1+\beta;\cos^2(\theta/2)) \qquad (7.14)$$

To find the symmetries we assume that after transformation we obtain the same expression but with the new indices.

We deduce that the new indices $n', \alpha', \beta', \rho', \sigma'$ in terms of the old one are:

$$n' = n + \alpha, \alpha' = -\alpha, \beta' = \beta, \rho' = \rho - \alpha, \sigma' = \sigma \tag{7.15}$$

In our case we find the Regge symmetry.

$$\begin{pmatrix} j_1 & j_2 & j_3 \\ m_1 & m_2 & m_3 \end{pmatrix} = \begin{pmatrix} (j_1+j_2-m_3)/2 & (j_1+j_2+m_3)/2 & j_3 \\ (j_1-j_2+m_1-m_2)/2 & (j_1-j_2+m_1-m_2)/2 & (-j_1+j_2) \end{pmatrix} \tag{7.16}$$

we can also find the others symmetries by applying this transformation to the other expressions of Jacobi polynomial. So the derivation of symmetries using the (IR) is much simpler than the Vilenkin method [29]. We may also use the other transformations of $_2F_1$ to deduce the other symmetries [53-55].

## 8. The integral representation of the 6j symbols of SU (2)

The theory of angular momentum is undoubtedly is the most studied in physics because it is of great interest in all branches of physics and chemistry. But we find by analogy with the 3j symbols that the (IR) of 6j symbols it is not emphasized in the literature so we will fill this gap.

The expression of 6j is:

$$\begin{pmatrix} j_1 & j_2 & j_3 \\ m_1 & m_2 & m_3 \end{pmatrix} \begin{Bmatrix} j_1 & j_2 & j_3 \\ l_1 & l_2 & l_3 \end{Bmatrix} =$$

$$\sum_{\mu_1 \mu_2 \mu_3} (-1)^{l_1+l_2+l_3+\mu_1+\mu_2+\mu_3} \begin{pmatrix} j_1 & l_2 & l_3 \\ m_1 & \mu_2 & -\mu_3 \end{pmatrix} \begin{pmatrix} l_1 & j_2 & l_3 \\ -\mu_1 & m_2 & \mu_3 \end{pmatrix} \begin{pmatrix} l_1 & l_2 & j_3 \\ \mu_1 & -\mu_2 & m_3 \end{pmatrix} \tag{8.1}$$

we put

$$m_1 = j_1, m_2 = -j_2, m_3 = j_2 - j_1, \mu_1 = \mu_3 - j_2 \text{ and } \mu_2 = \mu_3 - j_1. \tag{8.2}$$

With $J = l_1 + j_2 + l_3$
We get

$$\begin{pmatrix} j_1 & j_2 & j_3 \\ j_1 & -j_2 & (j_2-j_1) \end{pmatrix} \begin{Bmatrix} j_1 & j_2 & j_3 \\ l_1 & l_2 & l_3 \end{Bmatrix} = \sum_{\mu_1 \mu_2 \mu_3} (-1)^{l_1+l_2+l_3+\mu_1+\mu_2+\mu_3} \times$$

$$\begin{pmatrix} j_1 & l_2 & l_3 \\ j_1 & \mu_2 & -\mu_3 \end{pmatrix} \begin{pmatrix} l_1 & j_2 & l_3 \\ -\mu_1 & -j_2 & \mu_3 \end{pmatrix} \begin{pmatrix} l_1 & l_2 & j_3 \\ \mu_1 & -\mu_2 & (j_2-j_1) \end{pmatrix} \tag{8.3}$$

We find from the expression (6.6) that:

$$\begin{pmatrix} j_1 & l_2 & l_3 \\ j_1 & \mu_2 & -\mu_3 \end{pmatrix} = (-1)^{-j_1+l_2-\mu_3} \Lambda(j_1,l_2,l_3) \left[ \frac{(j_1+l_2-\mu_3)!(l_3+\mu_3)!}{(-j_1+l_2+\mu_3)!(l_3-\mu_3)!} \right]^{\frac{1}{2}} \tag{8.4}$$

With
$$\Lambda(j_1, l_2, l_3) = \left[\frac{(2j_1)!(-j_1+l_2+l_1)!}{(j_1+l_2+l_3+1)!(j_1-l_2+l_3)!(j_1+l_2-l_3)!}\right]$$

And
$$\Delta(j_1, j_2, j_3) = (J+1)! \prod_{i=1}^{3}(J-2j_i)!$$

The second term of (8.3) is derived from the above expression by permutation. And using the expressions (6.6) and (8.3) we obtain an expression containing this summation:

$$\sum_{\mu_3} \frac{(l_3+\mu_3)!}{(l_3-\mu_3)!(l_1-j_2+\mu_3)!(l_2-j_1+\mu_3)!}(tg\frac{\theta}{2})^{2\mu_3} = \frac{(2l_3)!}{(l_1+l_3-j_2)!(l_2+l_3-j_1)!} \times$$
$$_2F_1(-l_1-l_3+j_2, -l_2-l_3+j_1; -2l_3; -tg^2(\theta/2)) \qquad (8.5)$$

We find (8.5) by setting $l_3 - \mu_3 = r$ and using the formula $(x-r)! = x!/((-1)^r(-x)_r)$.

Using the expression (8.5) then we identify it with the first member we obtain the integral representation of the 6j:

$$\begin{Bmatrix} j_1 & j_2 & j_3 \\ l_1 & l_2 & l_3 \end{Bmatrix} = (-1)^{l_1+l_2-j_1-j_2} \sqrt{\frac{(j_1+j_2+j_3+1)!(j_1+j_2-j_3)!(2l_1)!(2l_2)!}{\Delta(j_1,l_2,l_3)\Delta(j_2,l_1,l_3)}} \times$$
$$\int[\left(\cos\frac{\theta}{2}\sin\frac{\theta}{2}\right)^{l_1+l_2}\left(tg\frac{\theta}{2}\right)^{j_1+j_2} d^{j_3}_{(-j_1+j_2, -l_1+l_2)}(\theta) \times$$
$$_2F_1(-l_1-l_3+j_2, -l_2-l_3+j_1; -2l_3; -tg^2(\theta/2))]\sin\theta \, d\theta \qquad (8.6)$$

## 9. Generating function of the 3j symbols for multiplicity-free of SU (3)

To determine the generating function of the 3j symbols we need the basis of the representation of SU (3), its generating function and the invariants of SU (3) which are very easy to find it in this case. The basis of the representation of SU (3) was constructed by many authors[3,7] using the lowering and raising operators expressed in terms of creation and annihilation operators of harmonic oscillators [4]. We've built the generating function of this basis by the method of coupling in the Fock Bargmann space [1, 16]. The invariants for 3j symbols of multiplicity-free are functions of the powers of the elementary invariants of SU (3) and the normalization is feasible in this case [51].

A brief review of the derivation of the basis of SU (3) and we'll just give the expression of the generating function then we perform the calculations to obtain the generating function of 3j symbols. Finally from the development of generating function we find the new expression of these symbols.

### 9.1 The basis of the group SU (2) ⊂ SU (3)

Let $D_{[\lambda,\mu]}$ the space of homogeneous polynomials and $V^{\lambda\mu}_{(t,t0,y)}(z^1, z^2)$ is the orthogonal basis with:
$$z^{(1)} = (z_1^1, z_2^1, z_3^1), \ z^{(2)} = (z_1^2, z_2^2, z_3^2), \ z_i^j \in C \qquad (9.1)$$

The space is homogeneous then

$$T_{11}V_{(\alpha)}^{\lambda\mu} = (\lambda+\mu)V_{(\alpha)}^{\lambda\mu}, \qquad T_{22}V_{(\alpha)}^{\lambda\mu} = (\mu)V_{(\alpha)}^{\lambda\mu} \qquad (9.2)$$

With

$$T_{ij} = \sum_k z_k^i (\partial/\partial z_k^i) \qquad (9.3)$$

The vectors $V_{(t,t_0,y)}^{\lambda\mu}(z^1, z^2)$ are eigenfunctions of the Casimir operator of the second order $\vec{T}^2$, the projection of $\vec{T}$ on the z axis and the hypercharge Y. The eigenvalues of these operators are respectively t (t + 1), $t_0$ and the triple of the hypercharge quantum number y. The numbers $t, t_0$ are the isospin and the component of isospin on the z axis. We have:

$$YV_{(\alpha)}^{\lambda\mu} = yV_{(\alpha)}^{\lambda\mu}, \qquad T_z V_{(\alpha)}^{\lambda\mu} = t_0 V_{(\alpha)}^{\lambda\mu}$$

And

$$\vec{T}^2 V_{(\alpha)}^{\lambda\mu} = t(t+1)V_{(\alpha)}^{\lambda\mu}. \qquad (9.4)$$

the condition of Young tableau on $V_{(\alpha)}^{\lambda\mu}$ imposes the further condition:

$$T_{12}V_{(\alpha)}^{\lambda\mu}(z^1, z^2) = 0 \qquad (9.5)$$

### 9.2 The generating function of the basis of SU (2) $\subset$ SU (3)

we determine the generating function by the method of coupling [31, 47].
We write

$$G((x,y,u),z) = \exp[f.\vec{z}^{(1)} + \vec{g}.\vec{z}^{(12)}] = \sum_{\lambda\mu tt_0 y} \varphi_{(t,t_0,y)}^{(\lambda\mu)}(\vec{f},\vec{g}) V_{(t,t_0,y)}^{(\lambda\mu)}(z^{(1)}, z^{(12)}) \qquad (9.6)$$

With $\qquad \vec{f} = (x_1\xi, x_1\eta, x_2), \quad \vec{g} = (y_1\eta, -y_1\xi, y_2) \qquad (9.7)$

and $\qquad \vec{z}^{(i)} = (z_1^i, z_2^i, z_3^i), \quad \vec{z}^{(ij)} = \vec{z}^{(i)} \times \vec{z}^{(j)} \qquad (9.8)$

we have

$$\varphi_{(t,t_0,y)}^{(\lambda\mu)}(\vec{f},\vec{g}) = N[(\lambda\mu),(\alpha)](x_1^p x_2^{\lambda-p} y_1^{(\mu-q)} y_2^q)(\xi^{(t+t_0)}\eta^{(t-t_0)}) \qquad (9.9)$$

And

$$N[(\lambda\mu),(\alpha)] = (-1)^q \sqrt{\frac{(\mu+p+1)!(\mu+\lambda-q+1)!}{(\lambda+1)!(2t+1)\lambda![p!(\lambda-p)!q!(\mu-q)!(t+t_0)!(t-t_0)!}}$$

We have also:

$$\begin{aligned} y &= -(2\lambda+\mu)+3(p+q), & 0 \le p \le \lambda, \\ t &= \mu/2+(p-q)/2, & 0 \le q \le \mu, \\ t_0 &= t-r, & r = 0,1,...,2t. \end{aligned} \qquad (9.10)$$

## 9.3 The invariant of the 3j symbols for multiplicity-free $\mu_1 = \mu_2 = 0$.

The invariant of 3j symbols of SU (3) is given by:

$$h = \sum_{\alpha_i} \begin{pmatrix} \lambda_1 0 & \lambda_2 0 & \lambda_3 \mu_3 \\ (\alpha_1) & (\alpha_2) & (\alpha_3) \end{pmatrix} V_{(\alpha_1)}^{(\lambda_1 0)}(z^1,0) V_{(\alpha_2)}^{(\lambda_2 0)}(z^3,0) V_{(\alpha_3)}^{(\lambda_3 \mu_3)}(z^{(5)}, z^{(56)})^c \quad (9.11)$$

The conjugate vector $(V_{(\alpha)}^{\lambda\mu})^c$ is deduced from $V_{(\alpha)}^{\lambda\mu}$ by R- Conjugation [6, 7, and 51]:

$$\lambda \leftrightarrow \mu, \ p \to \mu - q, q \to \lambda - p$$

And

$$(V_{(\alpha)}^{\lambda\mu})^c = (-1)^{y/2-t_0} V_{(-\alpha)}^{\lambda\mu}, \ (-\alpha) = (-y, t, -t_0)$$

the invariant are functions of the elementary invariants:

$$\vec{z}^{(1)} \cdot (\vec{z}^{(3)} \times \vec{z}^{(5)}), \ \vec{z}^{(1)} \cdot \vec{z}^{(56)}, \ \vec{z}^{(3)} \cdot \vec{z}^{(56)} \quad (9.12)$$

consequently we write:

$$h = N(k_i) \frac{[\vec{z}^{(1)} \cdot (\vec{z}^{(3)} \times \vec{z}^{(5)})]^{k_1} (\vec{z}^{(3)} \cdot \vec{z}^{(56)})^{k_2} (\vec{z}^{(1)} \cdot \vec{z}^{(56)})^{k_3}}{k_1! \quad k_2! \quad k_3!} \quad (9.13)$$

We can determine the constant of normalization by our method [47] but it is more simpler in this case to do the direct calculation [51], we write:

$$k_2 + k_3 = \lambda_3, \ k_1 + k_3 = \lambda_1, \ k_i \geq 0,$$
$$k_1 = \mu_3, \ k_1 + k_2 = \lambda_2, \quad (9.14)$$

So we have the decomposition:

$$(\lambda_1, 0) \otimes (\lambda_2, 0) \to \sum_{\mu_3} (\lambda_3 = \lambda_1 + \lambda_2 - 2\mu_3, \mu_3) \quad (9.15)$$

the normalization is:

$$N(k_i) = \sqrt{\frac{2\mu_3!(\lambda_2 - \mu_3)!(\lambda_1 - \mu_3)!}{(\lambda_1 + \lambda_2 - \mu_3 + 2)!(\lambda_1 + \lambda_2 - \mu_3 + 1)!}} \quad (9.16)$$

## 9.4 The generating function of the 3j symbols for multiplicity-free $\mu_1 = \mu_2 = 0$.

We find the 3j symbols from the expression (9.11) as:

$$\begin{pmatrix} \lambda_1 0 & \lambda_2 0 & \lambda_3 \mu_3 \\ (\alpha_1) & (\alpha_2) & (\alpha_3) \end{pmatrix} = \langle h \| V_{(\alpha_1)}^{(\lambda_1 0)}(z^1,0) V_{(\alpha_2)}^{(\lambda_2 0)}(z^3,0) V_{(\alpha_3)}^{(\lambda_3 \mu_3)}(z^{(5)}, z^{(56)})^c \rangle \quad (9.17)$$

Multiplying this expression by $(\prod_i t_i^{ki})$ and using (9.6) we write:

$$G((f,g),t) = \int \{\exp[\vec{f}_1 \cdot \vec{z}^{(1)}] \exp[+\vec{f}_3 \cdot \vec{z}^{(3)}] \exp[\vec{f}_5 \cdot \vec{z}^{(5)} + \vec{g} \cdot \vec{z}^{(56)}] +$$

$$[t_1 \vec{z}^{(1)} \cdot (\vec{z}^{(3)} \times \vec{z}^{(5)}) + t_2 \vec{z}^{(1)} \cdot \vec{z}^{(56)} + t_3 \vec{z}^{(3)} \cdot \vec{z}^{(56)}] d\mu(z^{(1)}, z^{(3)}, z^5, z^6) =$$

$$\sum_{\alpha_i} \begin{pmatrix} \lambda_1 0 & \lambda_2 0 & \lambda_3 \mu_3 \\ (\alpha_1) & (\alpha_2) & (\alpha_3) \end{pmatrix} (N(k_i))^{-1} \varphi_{(\alpha_1)}^{(\lambda_1 0)}(f_1) \varphi_{(\alpha_2)}^{(\lambda_2 0)}(f_3) \varphi_{(\alpha_3)}^{(\lambda_3 \mu_3)}(f_5, g) (\prod_i t_i^{ki}) \quad (9.18)$$

In carrying out the integration over $\vec{z}^{(1)}, \vec{z}^{(3)}, \vec{z}^{(6)}$ we find that the quantity in brackets is written as:

$$\exp[\vec{f}_5 \cdot \vec{z}^{(5)} + t_1 \vec{\bar{z}}^{(5)} \cdot (\vec{f}_1 \times \vec{f}_3) + (\vec{g} \times \vec{z}^{(5)}) \cdot ((t_2 \vec{f}_3 + t_3 \vec{f}_1) \times \vec{\bar{z}}^{(5)})] \qquad (9.19)$$

Let $\vec{h} = (t_2 \vec{f}_3 + t_3 \vec{f}_1)$ and using the expression

$$\vec{\bar{z}}^{(5)} \times (\vec{g} \times \vec{z}^{(5)}) = \vec{g}(\vec{\bar{z}}^{(5)} \cdot \vec{z}^{(5)}) - \vec{z}^{(5)}(\vec{\bar{z}}^{(5)} \cdot \vec{g}) \qquad (9.20)$$

we find for the third term the result:

$$\exp[(\bar{z}^{(5)})^t \begin{pmatrix} h_1 g_1 + h_2 g_2 & -h_1 g_2 & -h_1 g_3 \\ -h_2 g_1 & h_1 g_1 + h_3 g_3 & -h_2 g_3 \\ -h_3 g_1 & -h_3 g_2 & h_2 g_2 + h_3 g_3 \end{pmatrix} (z^{(5)})] \qquad (9.21)$$

using the formula (5.13) we find the generating function:

$$G((f,g),t) = \frac{1}{(1-\vec{h} \cdot \vec{g})^2} \exp[\frac{t_1}{1-\vec{h} \cdot \vec{g}} (\vec{f}_5 \cdot (\vec{f}_1 \times \vec{f}_3))] \qquad (9.22)$$

**9.5 Expression of 3j symbols of SU (3).**

We observe that $\vec{h} \cdot \vec{g} = t_2 \vec{f}_3 \cdot \vec{g} + t_3 \vec{f}_1 \cdot \vec{g}$.

And

$$\vec{f}_1 \cdot \vec{g} = [-x_1^1 y_1^5 [u^5 u^1] + x_2^1 y_2^5], \quad \vec{f}_3 \cdot \vec{g} = [x_1^3 y_1^5 [u^3 u^5] + x_2^3 y_2^5], \qquad (9.23)$$

$$\vec{f}_5 \cdot (\vec{f}_1 \times \vec{f}_3) = [x_2^1 x_1^5 x_1^3 [u^3 u^5] + x_1^3 x_2^5 x_1^1 [u^1 u^3] + x_1^5 x_1^1 x_2^3 [u^5 u^1]] \qquad (9.24)$$

developing (9.22)first and after that we use (9.23), (9.24) and the generating function (6.6) of the 3j symbols of SU(2) we find the expression of 3j symbols of SU (3):

$$\begin{pmatrix} \lambda_1 0 & \lambda_2 0 & \lambda_3 \mu_3 \\ (\alpha_1) & (\alpha_2) & (\alpha_3) \end{pmatrix} = \frac{(-1)^{y_3/2-(t_0)_3}(-1)^{(2t_1-\mu_3)} N(k_i)}{N[(\lambda_1 0),(y_1 00)]N[(\lambda_2 0),(y_2 00)]N[(\lambda_3 \mu_3),(\alpha_3)]} \times$$

$$\left[\sum_{k_1} (-1)^{k_1} \frac{1}{k_1!(\mu_3 - T + 2t_3 - k_1)!(T - 2t_1 - k_1)!(2t_1 - \mu_3 + k_1)!} \times \right.$$

$$\left. \frac{1}{(\lambda_2 - \mu_3 - T + 2t_1 + k_1)!(\lambda_1 - 2t_1 - k_1)!} \right] \times \begin{pmatrix} T_1 & T_2 & T_3 \\ (t_1)_0 & (t_2)_0 & (t_3)_0 \end{pmatrix}. \qquad (9.25)$$

We write the quantity between brackets:

$$[] = \frac{1}{(\mu_3 - T + 2t_3)!(T - 2t_1)!(2t_1 - \mu_3)!(\lambda_2 - \mu_3 - T + 2t_1)!(\lambda_1 - 2t_1)!} \times$$

$$_3F_2(-\mu_3 + T - 2t_3, -T + 2t_1, -\lambda_1 + 2t_1; 2t_1 - \mu_3 + 1, \lambda_2 - \mu_3 - T + 2t_1 + 1; 1) \qquad (9.26)$$

Thus the isoscalar factor expression is found for the first time in this compact form And which shows the great interest of our method.

**Acknowledgments**

   I thank Professor M. Kibler from the IPN-Lyon for his encouragements. I am also grateful to A.M. Rouquet for carefully reading my works written in French.

## 10. References

[1] J. Schwinger, In "Quantum Theory of angular momentum"
     Ed. L.C. Biedenharn and H. Van Dam, Acad. Press, NY, (1965)
[2] V. Bargmann and M. Moshinsky, Nuclear physics 18(1960) 697
[3] M. J. Moshinsky, Rev. Mod. Phys. 34(1962)813
[4] J. Nagel and M. Moshinsky, J. math. Phys. 6(1965)682
[5] G.E. Baird and L.C. Biedenharn, J. math. Phys. 4(1963)1449
[6] J.D. Louck Am. J. Phys. 38, 3, 1970
[7] W.J. Holman III and L.C. Biedenharn, "Group Theory and its Applications"(1971),
     Ed. Loebl (New York: Academic Press)
[8] A. O. Barut and R. Raczka, Theory of group representations
     PWN- Warszawa (1980).
[9] L. I. Schiff, "Quantum Mechanics", Third Edition. New York: McGraw-Hill, 1968.
[10] D. S. Saxon, "Elementary Quantum Mechanics" San Francisco: Holden-Day, 1968.
[11] E. Merzbacher, "Quantum Mechanics", Third Edition. New York: Wiley, 1998.
[12] A. Messiah," Mécanique quantique 1-2" E. Dunod, Paris (1965)
[13] L. D. Landau and E. M. Lifshitz. "Quantum Mechanics: Non-relativistic Theory"
     Pergamon Press, Oxford, 1977.
[14] J. J. Sakurai, "Modern Quantum Mechanics" (Addison Wesley 1994)
 [15] V. Bargmann, Commun. Pure Appl. Math. 14, 187 (1961).
[16] V. Bargmann, "On the representations of the Rotation group"
      Rev. Mod. Phys.34, 4, (1962)
[17] M. Hage Hassan and M. Lambert, Nucl. Phy. A188 (972) 545
[18] E.G. Lanza, M.V. Andrés, F. Catara, Ph. Chomaz, M. Fallot and J.A. Scarpaci
       Nucl. Phys.A788 (2007)112
[19] M. Hage-Hassan, "Generalization of Cramer's rule and its application to
      The projection of Hartree-Fock wave function" math-ph/arXiv: 0909.0129
[20] P. Ring and P.Schuck, "The nuclear many-body problem"
     3$^{rd}$ printing edition (2005) (Springer, Berlin, 1980).
[21] R. P. Feynman and A. R. Hibbs. "Quantum Mechanics and Path Integrals".
     McGraw-Hill, New York, 1965.
[22] D. S. Saxon, "Elementary Quantum Mechanics". San Francisco: Holden-Day, 1968.
[23] B. R. Holstein, "The harmonic oscillator propagator" Amer. J. Phys., 67 (1998)
[24] F.A.Barone, H. C. Boschi-Filho, and Farina,' Three methods for calculating the
      Feynman propagator' Am. J. Phys., Vol. 71, No. 5, May 2003, 490
[25] I. Erdelyi,"Heigher transcendal functions "Vol. 2, Mac.Graw-Hill, New-york (1953)
[26] M. Moshinsky, E. Sadurn´ı and A. del Campo, SIGMA 3 (2007), 110
[27] M. Hage-Hassan, "On the hydrogen wave function in Momentum-space, Clifford
     Algebra and the Generating function of Gegenbauer polynomials",
     ccsd-00305330, arXiv: math-ph/0807.4070

[28] M. Hage-Hassan, "The two-dimensional hydrogen atom in the momentum
     Representation", ccsd-00333701, arXiv: math-ph/0807.4070.
[29] N. J. Vilenkin, « Fonctions spéciales et théorie de la représentation des groupes »
     Dunod (1991).
[30] D.J. Rowe and C. Bahri, J. Math. Phys. 41, 6544 (2000)
[31] M. Hage-Hassan, J. Phys. A, 16 (1983)1835
[32] M. Hage-Hassan, "Angular momentum and the polar basis of harmonic oscillator"
     arXiv: [math-ph] /0903.0708
[33] Ashok Das, "Field Theory: A Path Integral Approach" (World Scientific,
     Singapore, 1993).
[34] Claude Itzyckson and Jean-Bernard Zuber, "Quantum Field Theory"
     (McGraw–Hill, New York, 1980).
[35] C. Grosche and F.Steiner," Classification of solvable Feynman path integrals"
     arXiv: [math-ph] /9302053
[36] P.Kutaanheimo and E. Steifel; J. Reine Angew. Math. 218, 204 (1965)
[37] M. Hage Hassan; "Inertia tensor and cross product in n-dimensions space"
     Preprint math-ph/0604051
[38] T. Levi-Civita opera Mathematiche (Bolognia), Vol.2 (1956)
[39] M. Boiteux, C. R. Acad. Scie. Série B274 (1972) 867.
[40] M. Hage Hassan, "Hurwitz's matrices, Cayley transformation and the Cartan-Weyl
     basis for the orthogonal groups", Preprint math-ph/0610021v1
[41] G.E. Baird and L.C. Biedenharn, J. math. Phys. 5(1964)1723
[42] A. R. Edmonds, "Angular Momentum in Quantum Mechanics"
     Princeton, U.P., Princeton, N.J., (1957)
[43] M. Hage-Hassan, J. Phys. A, 41, 9(1983)2891
[44] M.A.B.Beg and H.Ruegg, J.Math.Phys.6, 677(1965)
[45] T.J. Nelson, J.Math.Phys.8, 4(1967)857
[46] B.L. van der Waerden, Die gruppentheoreotische Methode in der
     Quantenmechanik(Julius Springer-Verlag, Berlin, (1932))
[47] M. Hage-Hassan, "On the Euler angles for the classical groups, Schwinger approach
     and Isoscalar factors for SU (3)" Preprint math-ph/0805.2740v1, 2008
[48] H. Weyl, Math.Z. 23, 271(1925)
[49] H. Weyl, Math.Z. 24, 377(1926)
[50] C. K. Chew and R. T. Sharp, Can. J. Phys., 44 (1966) 2789
[51] M. Resnikoff, J. math. Phys. 8(1967)63
[52] T. Regge," Symmetry properties of Clebsh-Gordan Coefficients "
     IL Nu.Cimento, 3 (1958)
[53] A.O.Barut and R. Wilson J. Math. Physics 17,6(1976)900
[54] J. Raynal, J. Math. Phys. 19 (1978), 467-476
[55] J. Raynal, "On the definition and properties of generalized 6-j symbols",
     J. Math. Phys. 20(1979), 2398-2415.